\documentclass[aps,prb,onecolumn,nofootinbib,citeautoscript,10pt]{revtex4-2}  
\usepackage{amsmath,amssymb,bm, makecell} 
\usepackage{graphicx}

\usepackage[tight]{subfigure} 

\usepackage{color} 
\usepackage[papersize={8.5in,11in}]{geometry}
\usepackage[colorlinks=true]{hyperref}
\hypersetup{        
    unicode=false,          
    pdftoolbar=true,        
    pdfmenubar=true,        
    pdffitwindow=false,     
    pdfstartview={FitH},    
    pdfkeywords={keyword1} {key2} {key3}, 
    pdfnewwindow=true,      
    colorlinks=true,       
    linkcolor=magenta, 
    citecolor=blue,        
    filecolor=magenta,      
    urlcolor=blue           
} 

\usepackage{graphicx}
\usepackage{dcolumn}
\usepackage{color}
\usepackage{amssymb,amsmath}
\usepackage{tabularx, graphicx}
\usepackage{latexsym}
\usepackage{colortbl}
\usepackage{psfrag}
\usepackage{bbm,bm,array,physics}
\usepackage{dsfont}

\makeatletter
\def\l@subsubsection#1#2{}
\makeatother

\def \nn{\nonumber \\}
 
 \def\*#1{\mathbf{#1}} 

\begin{document}
\title{Direction-dependent conductivity in planar Hall set-ups with tilted Weyl/multi-Weyl semimetals}

\author{Rahul Ghosh$^{1}$}
\author{Ipsita Mandal$^{1,2}$}
\email{ipsita.mandal@snu.edu.in}

\affiliation{$^1$Department of Physics, Shiv Nadar Institution of Eminence (SNIoE), Gautam Buddha Nagar, Uttar Pradesh 201314, India}

\affiliation{$^2$
Freiburg Institute for Advanced Studies (FRIAS), University of Freiburg, D-79104 Freiburg, Germany}

\begin{abstract} 
We compute the magnetoelectric conductivity tensors in planar Hall set-ups, which are built with tilted Weyl semimetals (WSMs) and multi-Weyl semimetals (mWSMs), considering all possible relative orientations of the electromagnetic fields ($\mathbf E $ and $\mathbf B $) and the direction of the tilt. The non-Drude part of the response arises from a nonzero Berry curvature in the vicinity of the WSM/mWSM node under consideration. Only in the presence of a nonzero tilt do we find linear-in-$ | \mathbf B| $ terms in set-ups where the tilt-axis is not perpendicular to the plane spanned by $\mathbf E $ and $ \mathbf B $. The advantage of the emergence of the linear-in-$ | \mathbf B| $ terms is that, unlike the various $| \mathbf B|^2 $-dependent terms that can contribute to experimental observations, they have purely a topological origin, and they dominate the overall response-characteristics in the realistic parameter regimes. The important signatures of these terms are that they (1) change the periodicity of the response from $\pi $ to $2\pi$, when we consider their dependence on the angle $\theta $ between $\mathbf E $ and $\mathbf B $; and (2) lead to an overall change in sign of the conductivity depending on $\theta$, when measured with respect to the $\mathbf B =0$ case.
\end{abstract}

\maketitle

\tableofcontents

\section{Introduction}

Aided by a combination of unprecedented advances in materials fabrication and theoretical analysis, the past decade has witnessed an explosive increase in the study of electronic systems having nodal points in the Brillouin zone (BZ), where two or more bands cross \cite{armitage_review, polash21_topological}. They are called semimetals because their energy bands are neither characteristic of those in insulators (as there is no gap) nor that of conventional metals (since the density of states vanish at the nodal points). In the category of three-dimensional (3d) semimetals, the most famous examples are the Weyl semimetals (WSMs) \cite{burkov11_weyl,yan17_topological}, whose energy dispersion relation in the vicinity of a band-crossing point is linear-in-momentum, resembling the relativistic Weyl equation (modulo the lack of strict Lorentz invariance). They are modelled by two-band Hamiltonians and host pseudospin-1/2 quasiparticles. Intriguingly, several of the defining physical properties of the Weyl fermions, such as the signature chiral anomaly (existing for odd
spatial dimensions), explained by Adler-Bell-Jackiw \cite{adler,bell}, continue to hold in these nonrelativistic settings involving condensed matter systems \cite{chiral_ABJ}.

A simple generalization of the WSM is a multi-Weyl semimetal (mWSM) \cite{bernevig,bernevig2,dantas18_magnetotransport}, whose dispersion is linear along one direction (which we choose to be the $z$-direction, without any loss of generality) and quadratic/cubic in the plane perpendicular to it (which we label as the $xy$-plane).
All these 3d nodal phases exhibit nontrivial topological features in the momentum space, because each nodal point acts as a source or sink of the Berry flux, which arises from the Berry phases. Because the Berry curvature is the analogue of the magnetic field, with the Berry connection acting as a vector potential, a nodal point can be thought of as a Berry monopole carrying integer units of charge. This is the same as the Chern number, a topological invariant, obtained by integrating the Berry curvature over a closed two-dimensional (2d) surface enclosing the nodal-point. Nielsen and Ninomiya's no-go theorem \cite{nielsen81_no} tells us that the nodes come in pairs in a given 3d BZ, such that each pair harbours Chern numbers $\pm J$ (where $J> 1$), acting as source and sink of the Berry flux and adding up to zero when summed over the two nodes. Intuitively, this satisfies the requirement that the Berry curvature flux lines must begin and end somewhere within the 3d BZ, which is obtained by imposing periodic boundary conditions on the real space lattice and Fourier transforming it.

The sign of the topological charge gives us the chirality $\chi $ of the associated node, leading to the nomenclature of right-moving and left-moving quasiparticles, corresponding to $\chi =1 $ and $\chi =-1 $, respectively. The values of the magnitude $J$ of the monopole charge at a Weyl (e.g., TaAs \cite{huang15_observation, lv_Weyl, yang_Weyl} and HgTe-class materials~\cite{ruan_Weyl}), double-Weyl (e.g., $\mathrm{HgCr_2Se_4}$~\cite{Gang2011} and $\mathrm{SrSi_2}$~\cite{hasan_mweyl16, singh18_tunable}), and triple-Weyl node (e.g., transition-metal monochalcogenides~\cite{liu2017predicted}) are $1$, $ 2$, and $ 3$, respectively. The chiral anomaly, mentioned in the beginning, refers to the phenomenon of charge pumping from one node to its conjugate in the presence of both electric ($\mathbf E$) and magnetic ($\mathbf B$) fields, which may be oriented in arbitrary directions relative to the separation of the pair of nodes. This leads to a local non-conservation of electric charge in the vicinity of an individual node, with the rate of change of the number density of chiral quasiparticles being proportional to $ J\,\mathbf E \cdot \mathbf  B $ \cite{chiral_ABJ, chiral_ano_mWSM}. 
However, on summing over the net chiral charges for the conjugate pairs of nodes in the entire BZ, the net value comes out to be zero, thus preserving the total charge. A direct consequence of the chiral anomaly is that for $\mathbf E \parallel \mathbf B $, the longitudinal conductivity along the applied magnetic field is proportional to $ B ^2 $ (where $B = |\mathbf B| $) and the intranode scattering time $\tau_{\rm inter}$, and its magnitude can be extremely large. Thus, the resistivity decreases with increasing magnetic field, leading to the observation of a large negative longitudinal magnetoresistance (LMR) in nodal-point semimetals~\cite{pallab_pixley_axial, lv21_experimental, huang15_observation, son13_chiral, seok_kim, moghaddam22_observation}.
However, recent investigations have shown that the interplay with the orbital magnetic moment (OMM) \cite{xiao_review, sundaram99_wavepacket} and strong internode scatterings can change the sign of the LMR within the semiclassical framework \cite{timm_pos_lmr}.

The chiral anomaly leads to another effect in topological nodal phases, namely the giant planar Hall effect (PHE) \cite{burkov17_giant, nandy_2017_chiral, das-agarwal_omm, nag21_magneto, onofre23_planar, ips_rahul_ph_strain}, which is the appearance of a large transverse voltage when $\mathbf B $ is not aligned along $\mathbf E $. Observation of the negative LMR and the PHE, with a specific dependence of the conductivity/resistivity tensors on the angle $\theta$ between $\mathbf E $ and $ \mathbf B $ (obtained theoretically from the semiclassical Boltzmann formalism), constitutes a telltale evidence for the chiral anomaly.\footnote{PHE is also observed in ferromagnets, but its value is very small.} Other smoking-gun signatures of nontrivial topology in bandstructures, studied widely in the literature, include intrinsic anomalous Hall effect~\cite{haldane04_berry,goswami13_axionic,burkov14_anomolous}, planar thermal Hall effect~\cite{son13_chiral, burkov17_giant,nandy_2017_chiral,nandy18_Berry, nandy19_planar,das19_linear, das-agarwal_omm, nag21_magneto, ips-serena}, magneto-optical conductivity when Landau levels need to be considered~\cite{gusynin06_magneto,staalhammar20_magneto,yadav23_magneto}, Magnus Hall effect~\cite{papaj_magnus,amit-magnus,sajid_magnus}, circular dichroism \cite{sekh22_circular,ips_cd}, circular photogalvanic effect \cite{moore18_optical,guo23_light,kozii, ips_cpge}, and transmission of quasiparticles across potential barriers/wells \cite{ips_aritra, ips-sandip, ips-sandip-sajid, ips_epj_trs}.

\begin{figure}[t]
\centering 
\subfigure[]{\includegraphics[width=0.2 \textwidth]{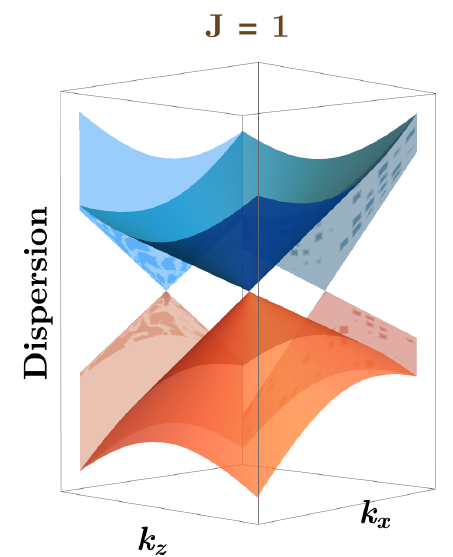}} \hspace{1 cm}
\subfigure[]{\includegraphics[width=0.2 \textwidth]{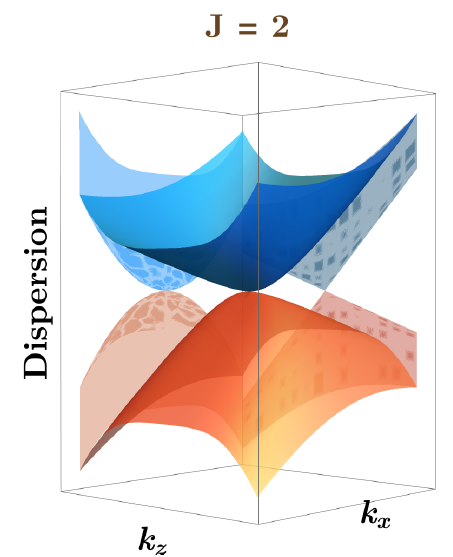}} \hspace{1 cm}
\subfigure[]{\includegraphics[width=0.2 \textwidth]{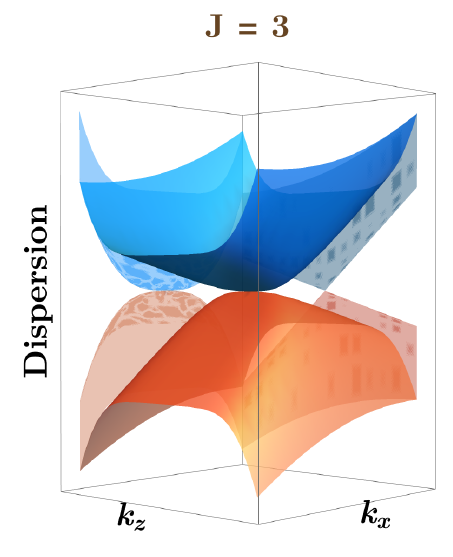}} 
\caption{
\label{fig_bands}
Schematic dispersion of a single tilted node [cf. Eq.~\eqref{eq_ham}] in a (a) Weyl, (b) double-Weyl, and (c) triple-Weyl semimetal, plotted against the $k_z k_x $-plane. The tilting is taken along the $k_z$-direction, along which the dispersion depends linearly on the momentum.
The double(triple)-Weyl node shows an anisotropic hybrid dispersion with a quadratic(cubic)-in-momentum dependence along the $k_x$-direction. The direction-dependent features are depicted more clearly with the help of the projections of the dispersion along the respective momentum axes.
}
\end{figure}

Anisotropy arising from tilting of the dispersion \cite{emil_tilted, trescher17_tilted} is often neglected because it enters the Hamiltonian with an identity matrix, thus not affecting the eigenspinors
and, hence, the topology of the low-energy theory in the vicinity of the band-crossing point (e.g., the Berry curvature of the Weyl cone does not depend on the tilt parameter). Tilting is generic in the absence of certain discrete symmetries, such as particle-hole and lattice point group symmetries. Therefore, tilted dispersion is expected to be present in WSMs/mWSMs in generic situations because of the generic nature of nodal points in 3d \cite{herring} (e.g., in a WSM with broken $ \mathcal T $ \cite{emil_tilted}), made possible by their topological stability.\footnote{The WSM/mWSM nodes are very robust, requiring only the discrete translational invariance of the crystal.} In the presence of a tilt for one of the momentum directions, which we take to be the $k_z$-component in this paper, the LMR has a term with linear-in-$ B $ behaviour \cite{zyuzin17_magnetotransport, sharma17_chiral}. Such linear terms may also arise when cubic-in-momentum band-bending terms ($\propto k_z^3 \, \sigma_z $) are included in the Hamiltonian \cite{cortijo16_linear}, which break the time-reversal symmetry $ \mathcal T $. Furthermore, in a PHE set-up, tilting leads to the presence of terms linearly dependent on $ B $ in the theoretical expressions of the longitudinal and transverse components of the magnetoelectric conductivity tensor \cite{ma19_planar, kundu20_magnetotransport, konye21_microscopic, shao22_plane}. This can explain the resistivity behaviour in experimental observations \cite{li18_giant}. Our aim is to demonstrate how tilt parameters as well as the intrinsic mixed linear-nonlinear dispersion of mWSMs lead to clear signatures in PHE, which are strongly direction-dependent. We show that the $ \mathcal T $-breaking induced by the tilt of nodes produces linear-in-$ B $ corrections, which depend on the direction of the tilt in a PHE set-up.

In this paper, we will consider an experimental set-up when a 3d semimetal is subjected to the combined effects of a static uniform external electric field $ \mathbf E $, applied along the unit vector
$\hat{\mathbf e}_E$, and a uniform external magnetic field $ \mathbf B $, applied along the unit vector $\hat{\mathbf e}_B $. If $\hat{\mathbf e}_E \cdot \hat{\mathbf e}_B \neq 0$, although the conventional Hall voltage induced from the Lorentz force is zero along $ \hat{\mathbf e}_B $, a node with a nonzero Berry monopole charge (i.e., a nonzero Chern number) gives rise to a voltage difference along this direction. This is the PHE discussed above~\cite{son13_chiral,burkov17_giant,li_nmr17,nandy_2017_chiral, nandy18_Berry,nag21_magneto, ips-serena}. The resulting components of the conductivity tensor, which lie in the $\hat{\mathbf e}_E\, \hat{\mathbf e}_B $-plane, are commonly known as the longitudinal magnetoconductivity (LMC) and the planar Hall conductivity (PHC). 
Their behaviour in various experimental set-ups has been extensively investigated in the literature~\cite{zhang16_linear, chen16_thermoelectric,das19_linear,das20_thermal,das22_nonlinear,pal22a_berry, pal22b_berry, fu22_thermoelectric,araki20_magnetic,mizuta14_contribution,ips_rahul_ph_strain}.
An untilted WSM is intrinsically isotropic and, hence, it will show the same response irrespective of how we choose to orient the $\hat{\mathbf e}_E$ and $\hat{\mathbf e}_B$ unit vectors. However, as soon as we introduce a tilt, it introduces an anisotropy, which should give rise to a direction-dependent response. For the mWSMs, even the untilted cases are anisotropic due to the fact that they feature a hybrid of a linear dispersion along one direction and a quadratic/cubic dispersion in the plane perpendicular to it.
Ref.~\cite{das19_linear} discusses the changes in the zero-temperature PHE response in WSMs, induced by changing the direction of $\mathbf B$ with respect to the tilt direction, demonstrating the emergence of a linear-in-$B$ term induced by a finite tilt. In another study \cite{pal22a_berry}, the authors have derived the response in such PHE set-ups using pseudospin-1 quasiparticles, described by three-band Hamiltonians, which have anisotropic hybrid dispersions analogous to the mWSMs. They did not include tilt in their computations and, hence, did not obtain any linear-in-$B$ term like we do in our current studies.


\begin{figure}[t]
\centering 
\subfigure[]{\includegraphics[width=0.25 \textwidth]{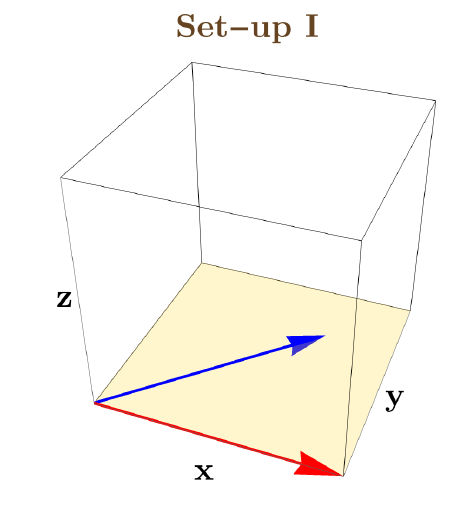}} \hspace{1 cm}
\subfigure[]{\includegraphics[width=0.25\textwidth]{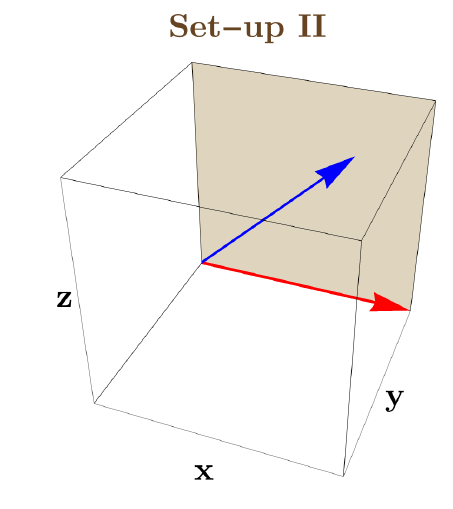}} \hspace{1 cm}
\subfigure[]{\includegraphics[width=0.25 \textwidth]{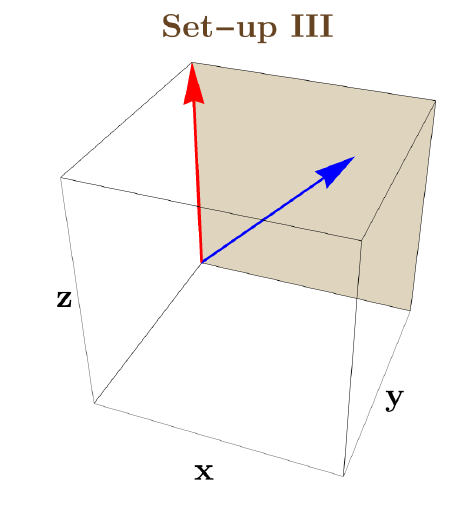}} 
\caption{Schematics of the three set-ups that we have used for investigating the planar Hall effect in WSMs/mWSMs, showing the relative orientations of the external homogeneous electric $\mathbf E $ (red arrow) and magnetic $ \mathbf B$ (blue arrow) fields, which we label as (a) set-up I, (b) set-up II, and (c) set-up III, respectively. The plane containing the $\mathbf E $ and $\mathbf B $ vectors (making an angle $\theta$ with each other) in each set-up has been highlighted by a background colour-shading. Each type of semimetal has a direction along which the dispersion is linear-in-momentum, which we have chosen to be the $z$-direction, which is also the axis with respect to which the dispersion has a tilt [cf. Fig.~\ref{fig_bands} and Eq.~\eqref{eq_ham}].  
\label{fig_setups}}
\end{figure}

Here we choose a tilt along the $k_z$-direction for each case, which maintains an isotropy in the $xy$-plane. The resulting dispersion is shown schematically in Fig.~\ref{fig_bands}. With these considerations in mind, we have three distinct configurations for applying $\mathbf E $ and $\mathbf B $ in a planar Hall set-up, which are illustrated in Fig.~\ref{fig_setups}. 
In the first two set-ups, which we label as I and II, $\hat{\mathbf e}_E$ is set perpendicular to the $k_z$-axis, which we have chosen to be the $x$-axis. In set-up I, we orient $\hat{\mathbf e}_B$ to lie in the $xy$-plane, while in set-up II, we orient $\hat{\mathbf e}_B$ to lie in the $zx$-plane. In the last set-up, which we denote as set-up III, $\hat{\mathbf e}_E$ is set parallel to the tilt-axis and $ \hat{\mathbf e}_B $ lies in the $zx$-plane. In each case, $\hat{\mathbf e}_B$ makes an angle $\theta $ with $\hat{\mathbf e}_A $, which is not $\pi/2$ in general for observing the PHE. We will employ the semiclassical Boltzmann transport formalism, which works well for small magnetic fields and small cyclotron frequency $\omega_c=e\,B/(m^*\, c) $ [where $m^* $ is the effective mass with the magnitude $\sim 0.11 \, m_e$ \cite{params2}, with $m_e$ denoting the electron mass] such that the Landau level quantization can be ignored. The regime of validity is given by $\hbar \, \omega_c \ll \mu$, where $\mu$ is the chemical potential/Fermi level.

The paper is organized as follows: In Sec.~\ref{sec_model}, we introduce the low-energy effective Hamiltonian in the vicinity of a tilted WSM/mWSM node. In Sec.~\ref{sec_cond}, we outline the steps to compute the conductivity tensor. Sec.~\ref{secres} is devoted to demonstrating the explicit expressions for the longitudinal and transverse components of the conductivity tensor, and illustrating their behaviour in some relevant parameter regimes. Finally, we conclude with a summary and outlook in Sec.~\ref{secsum}. 
In all expressions that follow, we will use the natural units by setting the reduced Planck's constant ($\hbar $), the speed of light ($c$), and the Boltzmann constant ($k_B $) to unity.
We also show the representative values of the various parameters, appearing in the Hamiltonians and the expressions for the conductivity tensors, using both the SI and the natural units, in Table~\ref{table_params}.

\section{Model for a tilted WSM/mWSM node}
\label{sec_model}

In the vicinity of a nodal point with chirality $\chi$ and Berry monopole charge of magnitude $J$, the effective continuum Hamiltonian is given by \cite{liu2017predicted,bernevig,bernevig2}
\begin{align} 
\label{eq_ham}
\mathcal{H}_J (  {\mathbf k}) & = 
\mathbf  d(  {\mathbf k}) \cdot \boldsymbol{\sigma}
+ \eta\,v_0\, k_z \, \sigma_0\, ,\quad
\mathbf d( \mathbf k) = \left \lbrace
\alpha_J \, k_\perp^J \cos(J \,\phi), \, \alpha_J \, k_\perp^J \sin(J \,\phi), \,
\chi \, v_0 \, k_z \right \rbrace,\nn
k_\perp & =\sqrt{k_x^2 + k_y^2}\,, \quad
\phi = \arctan({\frac{k_y}{k_x}})\,,
\quad \alpha_J =\frac{v_\perp}{k_0^{J-1}} \,, 
\end{align}
where $\chi = \pm 1$, $v_0$($v_\perp$) is the Fermi velocity along the $z$-direction($xy$-plane), $k_0$ is a material-dependent parameter with the dimension of momentum, and $\eta $ is the tilt parameter (which is taken along the $k_z$-direction). Henceforth, we will consider the positive-chirality node by setting $\chi = 1 $.
The eigenvalues of the Hamiltonian are given by
\begin{align} 
\label{eq_evals}
\mathcal{E}_{ \eta, s} ({  {\mathbf k}})= 
\eta\, v_0\, k_z  - (-1)^{s} \, |\mathbf d(  {\mathbf k})| \,, \quad
s \in \lbrace 1,2 \rbrace ,
\end{align}
where the value $1$($2$) for $s$ represents the upper(lower) band. The Hamiltonian for a WSM node is isotropic in absence of the tilt, which is recovered from $\mathcal{H}_J $ by setting $J=1$ and $\alpha_1= v_0$. Fig.~\ref{fig_bands} shows the schematic dispersion against the $k_z k_x$-plane.
In this paper, we restrict ourselves to the type-I phases such that $| \eta | < 1 $, which gives a Fermi point, an electron-pocket, or a hole-pocket, depending on whether the chemical potential cuts the nodal point, the upper band, or the lower band.

The Berry curvature (BC) associated with the $s^{\rm{th}}$ band is given by  \cite{xiao07_valley,nandy18_Berry}
\begin{align} 
\label{eqomm}
 {\mathbf  \Omega}_{ s}(  {\mathbf k}) & = 
    i \, \langle   \nabla_{  {\mathbf k}} \psi_s({  {\mathbf k}})| \, \cross  \, |  \nabla_{  {\mathbf k}} \psi_s({  {\mathbf k}})\rangle
\Rightarrow
\Omega^a_{ s}(  {\mathbf k})  =
\frac{-(-1)^s \,  \, \epsilon^{a}_{\,\, bc}
}
 {4\,| \* d( {\mathbf k}) |^3} \, 
 \* d( {\mathbf k}) \cdot
 \left[   \partial_{k_b} \* d( {\mathbf k}) \cross  \partial_{k_c} \* d( {\mathbf k}) \right ] ,
\end{align}
where the set of indices $ \lbrace a,b, c \rbrace $ takes values from $ \lbrace x, y, z \rbrace $, and is used to denote the components of the Cartesian vectors and tensors. The BC arises from the Berry phases generated by $|\psi_s (\mathbf k)\rangle$, where $\lbrace |\psi_s (\mathbf k)\rangle \rbrace $ denotes the set of orthonormal Bloch cell eigenstates for the single-particle Hamiltonian $\mathcal{H}_J (\mathbf k)$. On evaluating the expressions in Eq.~\eqref{eqomm} using Eq.~\eqref{eq_ham}, we get
\begin{align}
      {\mathbf  \Omega}_{ s }({  {\mathbf k}})= 
\frac{ (-1)^s \,
J \,v_0 \, \alpha_J^2 \, k_\perp^{2J-2} }
{2 \,\left | \mathbf  d(  {\mathbf k}) \right |^3
} \left
\lbrace k_x, \, k_y, \, J\, k_z \right \rbrace .
\end{align}
The Bloch velocity vector for the quasiparticles occupying the $s^{\rm th}$ is given by
\begin{align}
     {\boldsymbol{v}}_{ s } (  {\mathbf k}) = 
     \nabla_{  {\mathbf k}} \,  \mathcal{E}_{\eta, s}  ({  {\mathbf k}}) 
  = - \frac{ (-1)^s}  { \left | \mathbf  d(  {\mathbf k}) \right |} 
\left \lbrace J \, \alpha_J^2 \, k_\perp^{2J-2} \, k_x, \, J \, 
		\alpha_J^2 \, k_\perp^{2J-2} \, k_y, \, v_0^2 \, k_z
		\right \rbrace     
  + \Big\lbrace 0, 0, v_0 \,\eta \Big\rbrace.
\end{align}
We find that the BC is independent of the tilt, as expected, while the $z$-component of $ {\boldsymbol{v}}_{ s } (\mathbf k )$ gets shifted due to the tilt term. 
In this paper, we will take a positive value of the chemical potential $\mu$ such that it cuts the conduction band with $s=1$. Henceforth, we use the notations $ \mathcal{E}_{\eta, 1}  = \mathcal{E} $, $ {\mathbf  \Omega}_{ 1 } =  {\mathbf  \Omega}_{}$, and $ {\boldsymbol{v}}_1 =  {\boldsymbol{v}} $, in order to avoid cluttering.

\section{Magnetoelectric conductivity}
\label{sec_cond}

We use the semiclassical Boltzmann formalism to find the form of the magnetoelectric conductivity tensor
$\sigma_{p q }$ in a generic PHE set-up. We employ a relaxation-time-approximation for the collision integral and, furthermore, assume a momentum-independent relaxation time $\tau $, which we treat as a phenomenological constant. We focus on the scenario when the internode scattering amplitude is negligible compared to the intranode scattering amplitude, which suppresses any relaxation towards equal occupation of the two nodes in a conjugate pair, thereby enhancing signatures of the chiral anomaly \cite{son13_chiral, seok_kim}.
Therefore, $\tau $ corresponds solely to the intranode scattering time.
The derivation is outlined in detail in Refs.~\cite{ips-kush-review, ips_rahul_ph_strain}.
The final expression for conductivity, contributed by the conduction band at the node with positive chirality, is given by
\begin{align}
\label{eqsigmatot}
& \sigma_{ p q  }^{\rm tot} = \sigma_{ p q }^{\rm AHE} + \sigma_{ a b }^{ \mathbf \Gamma}
		                 + \sigma_{ p q } \,, \quad
\sigma_{ p q }^{\rm AHE} = -{e^2}  \,
		\int \frac{ d^3 \mathbf k}{(2\, \pi)^3 } \, \epsilon_{p q r} \,
\Omega^r \,  f_{eq}(\mathcal{E}) \,,\nn
&  \sigma_{ p q } =- e^2 \, \tau
\int \frac{ d^3 \mathbf k}{(2\, \pi)^3 } \, \mathcal{D} 
\left [ v_p  + {e\, B_p}  \left( 
{\boldsymbol{v}} \cdot \mathbf \Omega \right) \right ]
		\left [ v_q  + e\,  B_q    \left( 
{\boldsymbol{v}} \cdot \mathbf \Omega \right) \right ]
		\, \frac{\partial  f_{eq} (\mathcal{E}) } {\partial  \mathcal{E} } \,,
	\end{align}
where
\begin{align}
\mathcal D = \left( 1 +  {e} 
\, {\mathbf B} \cdot \mathbf{\Omega }  \right)^{-1} 
\end{align}
is the phase-space-factor, which takes into account the modified density of states in the presence of an external magnetic field, and
\begin{align}
f_{eq} \big (\epsilon  \big )
= \frac{1}
{e^{  \beta \,\left (\epsilon  -\mu  \right) } + 1} 
\end{align}
represents the equilibrium distribution function of the fermionic quasiparticles (with $\beta^{-1}  = k_B\, T $). We have taken the charge of each quasiparticle to be $-e$, where $e$ is the magnitude of the charge of an electron. The first part, labelled as $\sigma_{ p q }^{ \rm AHE}$, represents the ``intrinsic anomalous'' Hall effect \cite{haldane04_berry, goswami13_axionic, burkov14_anomolous}. The second part $ \sigma_{ p q }^{ \mathbf \Gamma}$ is the  Lorentz-force-contribution to the conductivity. The last term, $  \sigma_{ p q } $, is the Berry-curvature-related conductivity coefficient. For a momentum-independent $\tau$, $ \sigma_{ p q }^{ \mathbf \Gamma}$ is an order of magnitude smaller than the other terms~\cite{nandy_2017_chiral} in a typical WSMs/mWSM and, hence, we neglect it. Furthermore, we are not interested in $\sigma_{ p q }^{ \rm AHE}$ because, together with the contribution from the node with opposite chirality, it leads to an overall zero contribution when we sum the conductivity coming from the two nodes.

For the ease of calculations, we decompose $\sigma_{ p q}$ into four parts as follows:
\begin{align}  
\label{eq_sig_4parts}
 \sigma_{ p q}^{ (1) } & = \tau \, e^2 
  \int \frac{d^3   {\mathbf k}} {(2  \, \pi)^{3}} \, \mathcal{D} \,
   \,  v_p (  {\mathbf k}) \,  v_q(  {\mathbf k}) \,
  \left [ -\frac{\partial f_{eq} (\mathcal E)}   {\partial \mathcal E} \right ] , \nn
 \sigma_{ p q} ^{(2)}  & = 
 B_p \, B_q \, \tau \, e^4
\int \frac{d^3   {\mathbf k}}{(2  \, \pi)^{3}} \, \mathcal{D} 
 \, \left [  {\boldsymbol v}   (  {\mathbf k})  \cdot   {\mathbf  \Omega} (  {\mathbf k})  \right ]^2 
  \left [ -\frac{\partial f_{eq} (\mathcal E)}   {\partial \mathcal E} \right ] , \nn
 \sigma_{ p q} ^{(3)}  & =   
{B_q \, \tau \, e^3}   
 \int \frac{d^3  \mathbf  k}{(2  \, \pi)^{3}} \, \mathcal{D} \, 
  v_p (  {\mathbf k}) \left [  {\boldsymbol v}   (  {\mathbf k})  \cdot   {\mathbf  \Omega} (  {\mathbf k})  \right ] 
  \left [ -\frac{\partial f_{eq} (\mathcal E)}   {\partial \mathcal E} \right ] , \nn 
  \sigma_{ p q} ^{(4)}  & =  
{B_p \, \tau \, e^3  }
  \int \frac{d^3   {\mathbf k}}{(2  \, \pi)^{3}} \, \mathcal{D} 
  \,  v_q (  {\mathbf k}) \left [  {\boldsymbol v}   (  {\mathbf k})  \cdot   {\mathbf  \Omega} (  {\mathbf k})  \right ]  \left [ -\frac{\partial f_{eq} (\mathcal E)}   {\partial \mathcal E} \right ] .
\end{align}
We find that $\sigma_{ p q} ^{(2)} $, $\sigma_{ p q} ^{(3)} $, and $\sigma_{ p q} ^{(4)} $ go to zero if the BC vanishes. We will expand $\mathcal D $ in powers of $\mathbf B \cdot \mathbf \Omega $, restricting to the regime with a weak strength of the magnetic field, and keep terms upto order $ \left( \mathbf B \cdot \mathbf \Omega \right)^2 $ in the expressions for $\sigma_{ p q} $.

Now we outline the details for obtaining the final form of the LMC for the set-up I, discussed in Sec.~\ref{eqset1}. For this case, we have to set $ \hat{\mathbf e}_E = {\mathbf{\hat x}}  $ and $\hat{\mathbf e}_B = \cos \theta \, {\mathbf{\hat x}} + \sin \theta \, {\mathbf{\hat y}}  $. Hence, the LMC is given by
\begin{align}
 \sigma_{xx} = 
\sigma_{xx} ^{(1)}   + \sigma_{xx}^{(2)}   +   \sigma_{ xx } ^{(3)}   + \sigma_{ xx } ^{(4)}  \,,  \nonumber
\end{align}
where
\begin{align}
\label{eq_sxx_comp}
 \sigma_{xx} ^{(1)}  & = -
 \tau \, e^2  \int \frac{d^3 \mathbf k}{(2 \pi)^{3}} \, \mathcal{D} \, 
 v_x^2( {\mathbf  k}) 
\, \frac{\partial f_{eq} (\mathcal{E})
 }
 {\partial \mathcal{E}} \,,\quad
 \sigma_{xx}^{(2)}  = -
 \tau \, e^4 \, B_x^2  
 \int \frac{d^3  {\mathbf  k}}{(2 \pi)^{3}} \, 
 \mathcal{D} \left[  {\boldsymbol v}( {\mathbf  k}) \cdot \mathbf \Omega( {\mathbf  k}) \right ] ^2 
 \, \frac{\partial f_{eq} (\mathcal{E})
 }
 {\partial \mathcal{E}} \,  , \nn
\sigma_{ xx } ^{(3)}  & = \sigma_{ xx } ^{(4)}  = - \, \tau \, e^3 \, B_x 
\int \frac{d^3  {\mathbf  k}}{(2 \pi)^{3}} \, 
\mathcal{D} \, v_x( {\mathbf  k}) 
\left[{\boldsymbol v}( {\mathbf  k}) \cdot \mathbf \Omega( {\mathbf  k}) \right ] 
\,  \frac{\partial f_{eq} (\mathcal{E})
 }
 {\partial \mathcal{E}} \,   .
\end{align}

Keeping terms upto $\mathcal{O} \big( B^2 \big) $, we obtain
\begin{align}
\sigma_{xx} ^{(1)}  = - \frac{\tau \, e^2 }{(2 \pi)^{3}} \, 
\sum_{ j =0}^2 
\int_0^\infty dk_\perp \, \int_{-\infty}^{\infty} \, dk_z \, \int_0^{2\pi} \, d\phi \, 
k_{\perp}  \, 
v_x^2  \,\left[
- e \, ( { \mathbf B }  \cdot  {\mathbf  \Omega}) \right ]^j  
\frac{\partial f_{eq} (\mathcal{E}) }    {\partial \mathcal{E}} \,.
\end{align}
Changing the integration variables as
\begin{align}
k_\perp = \alpha_J^{-  1/J } \, \epsilon^{  1/J } 
 \left( \sin \gamma   \right)^{  1/J }\,,\quad
k_z = \frac{  \epsilon \,  \cos \gamma }
{v_0}  \,, \quad
k_x = k_\perp \cos \phi\,, \text{ and } k_y = k_\perp \sin \phi
\end{align}
leads to $\mathcal{E}_{\eta,s} = \epsilon \, \Theta_s(\gamma)$, where $\Theta_s(\gamma) = (-1)^{s+1} + \eta  \cos \gamma $ in general (i.e., irrespective of whether we consider the upper or the lower band). Since we are considering $s=1$ here, we have
\begin{align} 
\label{eq_sigmaxx1_app}
 \sigma_{xx} ^{(1)}    = 
 \frac{\tau \, e^2  }{(2 \pi)^{3}} \, \frac{J}{v_0} \,
 \sum_{j=0}^2 
\left (  \frac{e \, \, v_0 \, J \, \alpha_J^{  1/J }}{2} \right )^j 
\int_0^{2 \,\pi} d\phi 
\cos^2{\phi} \left[ B \,\cos{(\theta - \phi)}  \right]^j   
\int_{0}^{\pi} d\gamma \left (\sin{\gamma} \right )^{3 + j \, \left (2 -  1/J\right  )}
\, \mathcal{I}_1(j, \gamma)\,,
\end{align}   
where
\begin{align}
 \mathcal{I}_1(j, \gamma) = 
  \int_{0}^\infty  d\epsilon\,
  \epsilon^{2 - j\, \left (1 +   1/J \right )} 
  \frac{\beta \, e^{\beta \left (\mathcal{E}- \mu \right )}}
  { \left [ 1 + e^{\beta \, (\mathcal{E} - \mu)} \right ]^2}  \,.
\end{align}
Following a similar sequence of steps, the first part of the PHC is given by
\begin{align} 
 \sigma_{yx} ^{(1)}    = 
 \frac{\tau \, e^2  }{(2 \pi)^{3}} \, \frac{J}{v_0} \,
 \sum_{j=0}^2 
\left (  \frac{e \, \, v_0 \, J \, \alpha_J^{  1/J }}{2} \right )^j 
\int_0^{2 \,\pi} d\phi \sin \phi\,
\cos \phi \left[ B \,\cos{(\theta - \phi)}  \right]^j   
\int_{0}^{\pi} d\gamma \left (\sin{\gamma} \right )^{3 + j \, \left (2 -  1/J\right  )}
\, \mathcal{I}_1(j, \gamma)\,.
\end{align}

\begin{table}[]
\centering
\begin{tabular}{|c|c|c|}
		\hline
Parameter &   SI Units &   Natural Units  \\ \hline
$v_0$ from Refs.~\cite{params2} & $15\times10^{5} $ m~s$^{-1} $ & $0.005$  \\ \hline
$\tau$ from Refs.~\cite{params2, watzman18_dirac} & $ 10^{-13} \, \text{s} $ & $ 152 $ eV$^{-1}$  \\ \hline
$ T $ from Refs.~\cite{nag21_magneto, li18_giant} & $ 0 - 100 \, \text{K} $ & 
$0 -  8.617 \cross 10^{-3} \text{eV} $ \\ \hline
$ B $ from Ref.~\cite{ghosh20_chirality} &  $ 0 - 10  $ Tesla 
& $ 0 - 1950  $ eV$^{2}$
\\ \hline
$\mu$ from Refs.~\cite{pal22a_berry, sharma17_chiral} & $  1.6\times 10^{-21} -  1.6\times 10^{-20} $ J 
& $0.01 - 0.1$ eV \\ \hline
\end{tabular}
\caption{\label{table_params}
We list the parameter regimes which we have used in plotting the components of conductivity tensor.
In terms of natural units, we need to set $\hbar=c=k_{B}= e = 1$. In our plots, we have used $v_\perp = v_0$ (from the table entry), $\alpha_{2}=3.9 \times 10^{-5}$ eV$^{-1}$, and $\alpha_{3}=2.298 \times 10^{-6}$ eV$^{-2}$. 
While for $ J=2$ and $J=3$, $v_\perp$ has been set equal to $ v_0$ for the sake of simplicity,
the isotropic dispersion (modulo the tilt) of $J =1$ constrains us to have $v_{\perp} = v_0$.
}
\end{table}

In order to evaluate $\mathcal{I}_1(j, \gamma)$, we employ the Sommerfeld expansion
\begin{align}  
\label{epsilon integration}
&  \int_{0}^\infty d\epsilon \, \epsilon^n \,
\frac{\beta\, e^{\beta(\mathcal{E}- \mu)}}
{ \left [ 1 + e^{\beta \, \left (\mathcal{E}- \mu \right )} \right ] ^2}  
= \frac{1}
{\left[ \Theta_1 (\gamma) \right ]^{n+1}}
\int_{0}^\infty  d\mathcal{E} \, \mathcal{E}^n
 \frac{\beta \, e^{\beta\, (\mathcal{E} - \mu)}}
 {(1 + e^{\beta \, (\mathcal{E}- \mu)})^2} \
 = \frac{ \Upsilon_n ( \mu, \beta) }
 {\left[ \Theta_1(\gamma) \right ]^{n+1}} \,,\nn
& 
\Upsilon_n ( \mu, \beta) =
 \mu^n \left[
 1 + \frac{\pi^2 \, n\, (n-1)}  {6  \, \beta^2 \, \mu^2} 
 + \frac{7 \, \pi^2 \,n \,(n-3) \,(n-2) \,(n-1) }
 {360  \, \beta^4 \, \mu^4 } 
+ \order{\left(  {\beta \,\mu } \right)^{-6}}
 \right ],
\end{align}
which is valid in the regime $\beta \mu \gg 1$.
Plugging in the above expression, we get 
\begin{align}
\label{eqsxxpart1}
 \sigma_{xx} ^{(1)}    = 
 \frac{\tau \, e^2  }{(2 \pi)^{3}} \, \frac{J}{v_0} \, 
 \sum_{j=0}^2 
  \Upsilon_{2 - j\, (1 +   1/J )}( \mu, \beta) 
\left[  \frac{e \, \, v_0 \, J \, \alpha_J^{  1/J }}{2 } \right ]^j
 \,  \int_0^{2\pi} d\phi \cos^2{\phi} 
 \left[  B \,\cos{(\theta - \phi)}  \right]^j  
  \int_{0}^{\pi} d\gamma\,
   \frac{(\sin{\gamma})^{3 + j \,\left (2 -   1/J \right )}}
  { \left[ \Theta_1(\gamma) \right]^{3 - j\, (1 +   1/J )}} \,.
\end{align}  
The last step is to perform the $\gamma$-integral, using the identity
\begin{align}
& \int_{0}^\pi d\gamma\,
 \frac{ \left (\sin{\gamma} \right )^{m} \left (\cos{\gamma} \right)^n}
 {(1+\eta  \cos{\gamma})^l} 
= \int_{-1}^1 dx\,  
\frac{\left(1-x^2\right)^{\frac{m-1}{2}} x^n}{(1+\eta\,  x)^l}  \nn
 & = \frac{\sqrt{\pi} \, \, \Gamma (\frac{m+1}{2})}  
 {4} 
\Bigg [ 2 \,\left \lbrace  (-1)^n + 1 \right \rbrace 
  \,  \Gamma \Big(\frac{n+1}{2} \Big) \,
\,   _3\tilde{F}_2 \Big(\frac{n+1}{2},\frac{l+1}{2},\frac{l}{2};
    1/2,\frac{m+n+2}{2} ;\eta^2\Big )    \nn
& \hspace{2.75 cm} 
+ \eta  \, l
\left \lbrace  (-1)^n - 1 \right \rbrace \,
 \Gamma \Big(\frac{n}{2}+1\Big ) \, 
 _3\tilde{F}_2 \Big(\frac{n+2}{2},\frac{l+1}{2},\frac{l+2}{2}; 3/2 ,\frac{m+n+3}{2} ;\eta^2 \Big)
 \Bigg ]\,,
\end{align}
where $_{n_1}\tilde{F}_{n_2}\big( \{a_1,\ldots, a_{n_1} \};\{b_1,\ldots, b_{n_2} \}; X \big)$ is the regularized hypergeometric function \cite{hypergm}. 
Implementing this, we finally get
\begin{align} 
\label{eqsxxI1}
\sigma_{xx} ^{(1)}  &  = \sigma_{{\text{Drude}}} 
+\sigma_{xx}^{{(1, \text{BC})}}\,,\\
\label{eqdrude}
   \sigma_{{\text{Drude}}}
   &=   \,J \, e^2\, \tau\,\,  \Upsilon_{2} ( \mu, \beta )
 \left[
  \frac{\delta_{\eta,0} }
   {6 \,\pi^2 \, v_0}  
 +  \frac{  \delta_{\eta,0}-1}
 {4 \, \pi^2\, \eta^3 \,v_0} 
  \left(\frac{\eta }{\eta^2-1} + \tanh^{-1} \eta \right)  \right] , \\
\label{eqber}
 \sigma_{xx}^{{(1, \text{BC})}} &=     
 \frac{J^3 \,e^4  \,\tau  \,
B^2 \, \alpha^{{2}/{J}}  
\left(3 \,\cos^2 \theta + \sin^2 \theta \right)  
 \,\Gamma (4-  1/J ) \, v_0 
}
 {128 \,\pi^{ 3/2 } } 
\,  _2\tilde{F}_1  ( 1/2-  1/J ,(J-1)/J ; 9/2-  1/J ;\eta^2 ) 
 \,\Upsilon_{-{2} / {J}} (  \mu, \beta  )\,,
\end{align}
where $\sigma_{{\text{Drude}}} $ is independent of the magnetic field and is usually referred to as the Drude contribution. The Drude part arises from the $j=0$ term in the summation appearing in the integrand on the right-hand-side of Eq.~\eqref{eqsxxpart1}. The $j=1 $ term vanishes identically because of the presence of odd powers of $\cos{\phi}$ or $\sin \phi $. The $j=2 $ term gives rise to the second nonzero part $\sigma_{xx}^{{(1, \text{BC})}}$, whose origin is from a nonzero BC. For the case of PHC, we find that $j = 0$ and $j=1 $ terms vanish while performing the $\phi$-integration, and $j = 2$ provides the only nonzero contribution to $\sigma_{yx} ^{(1)}$, which has a purely topological origin.
Although we do not have any linear-in-$B$ term for this case, we will find it in the other set-ups. 
Proceeding in the way explained above, we obtain the final expressions for $\sigma_{xx}^{(2)} $, $\sigma_{ xx } ^{(3)} $, $\sigma_{ xx } ^{(4)} $, $\sigma_{yx}^{(2)} $, $\sigma_{ yx } ^{(3)} $, and $\sigma_{ yx } ^{(4)}$, the steps of which are not very necessary to write down explicitly.

To show the emergence of the linear-in-$B$, we include here the starting expression for the first part of the LMC of set-up II, discussed in Sec.~\ref{eqset2}. This is captured by
\begin{align}  
\sigma_{xx}^{(1)}   = 
\frac{\tau \, e^2  }{(2 \pi)^{3}} \, \frac{J}{v_0} \, 
\sum_{j=0}^2 
 \left ( \frac{e \,B\,  J \, \alpha_J^{1/J}} {2} \right )^j 
  \int_0^{2\pi} d\phi \,\cos^2{\phi} 
   \int_{0}^{\pi} d\gamma \,  
  \left  (\sin{\gamma} \right )^{3+ j \,\left (2-2/J \right)} 
  \, \mathcal{I}_2(j, \gamma, \phi) \,,
\end{align}  
where 
\begin{align}
\mathcal{I}_2(j, \gamma, \phi) =
 \int_{0}^\infty   d\epsilon \, 
 \epsilon^{2 - j\,\left  (1 + 2/J \right )}  
 \left[ 
 v_0  \cos{\theta}  \cos{\phi}  \,  
 (\epsilon \, \sin{\gamma})^{1/J} + J  \, \alpha^{1/J} \, \epsilon \sin \theta \cos \gamma
 \right]^j    
 \frac{\beta \, e^{\beta \left (\mathcal{E}- \mu \right )}}
  { \left [ 1 + e^{\beta \, (\mathcal{E} - \mu)} \right ]^2} \,.
\end{align}
For $j=0$, we get the Drude part, which is the same as Eq.~\eqref{eqdrude} obtained for set-up I. For $j=1$, we now have a nonvanishing contribution coming from the $\phi$-independent part of $\mathcal{I}_2(1, \gamma, \phi)$, because, unlike the LMC in set-up I, the integrand here has even powers of $\cos \phi$. This leads to a term linear-in-$B$. For $j=2$, we obtain the usual $B^2$-dependent terms, which exist even in the absence of a nonzero tilt.

The intermediate steps for the remaining tensors are omitted here, as they are similar to the derivations demonstrated above. In any case, we provide the full final expressions for each set-up in the corresponding subsections of Sec.~\ref{secres}.

\section{Results}
\label{secres}

In this section, we explicitly write down the expressions for the nonzero components of the magnetoelectric conductivity tensor for the three distinct set-ups shown in Fig.~\ref{fig_setups}. We also illustrate their behaviour by some representative plots, and discuss the physical implications of their characteristics. Before delving into the explicit expressions, let us first summarize the results in Table~\ref{table_results} for the ease of a quick comparison:
\begin{table}[h]
\centering
\begin{tabular}{ |c|c|c|c|}
\hline
    & Set-up I &  Set-up II &  Set-up III  \\ \hline
LMC & \makecell{terms proportional to $B^0$ and $B^2$} & \makecell{terms proportional to $B^0$, $B$, and $B^2$} & 
terms proportional to $B^0$, $B$, and $B^2$  \\ \hline
PHC &
\makecell{terms proportional to $B^2$} &
\makecell{terms proportional to $B$ and $B^2$} &
\makecell{terms proportional to $B$ and $B^2$}  \\ \hline
\end{tabular}
\caption{\label{table_results}
Summary of the key characteristics of the response for the three distinct set-ups, showing the $B$-dependence of the terms appearing in the final expressions.
}
\end{table}

\begin{figure}[t]
\centering
\subfigure[]{\includegraphics[width=0.99 \textwidth]{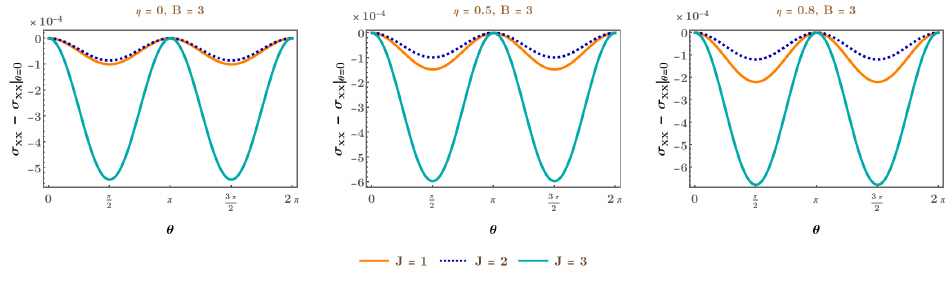}} 
\subfigure[]{\includegraphics[width=0.99 \textwidth]{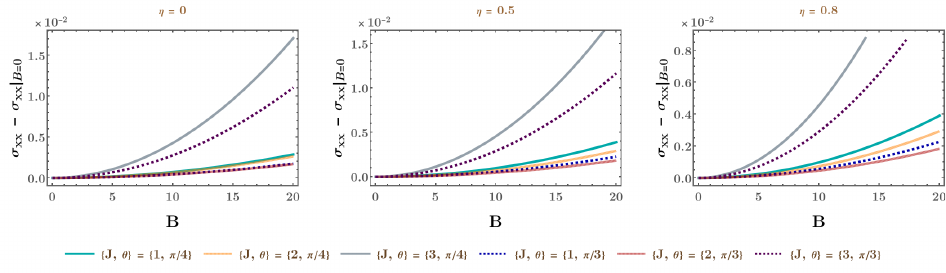}} 
\caption{Set-up I: Variation of the LMC (in units of eV) with (a) the angle $\theta$ and (b) the magnitude $ B $ (in units of eV$^2$) of the magnetic field. For all the plots, we have set $v_0 = 0.005$, $\tau=151$ eV$^{-1}$, $\mu = 0.1$ eV, and $\beta=1160$ eV$^{-1}$.
\label{fig_C1sigmaxx}}
\end{figure}

\begin{figure}[t]
\centering
\subfigure[]{\includegraphics[width=0.99 \textwidth]{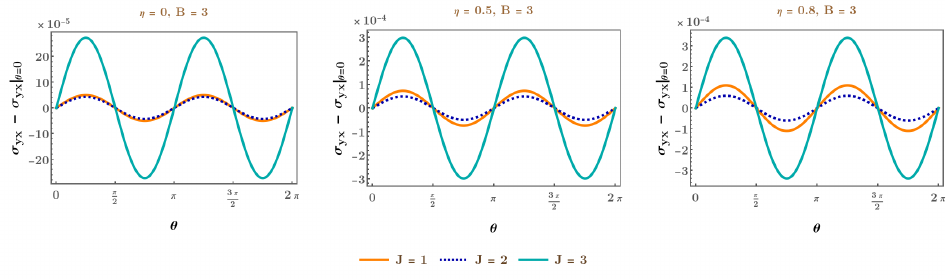}} 
\subfigure[]{\includegraphics[width=0.99 \textwidth]{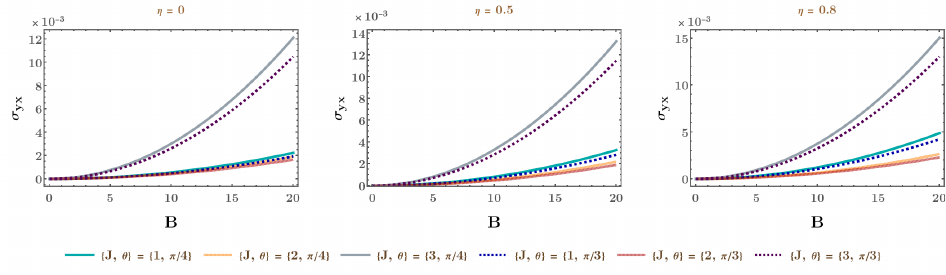}} 
\caption{Set-up I: Variation of the PHC (in units of eV) with (a) the angle $\theta$ and (b) the magnitude $ B $ (in units of eV$^2$) of the magnetic field. For all the plots, we have set $v_0 = 0.005$, $\tau=151$ eV$^{-1}$, $\mu = 0.1$ eV, and $\beta=1160$ eV$^{-1}$.
\label{fig_C1sigmayx}}
\end{figure}

\subsection{Set-up I}
\label{eqset1}

In set-up I, as shown in Fig.~\ref{fig_setups}(a), the tilt-axis is perpendicular to the plane spanned by $\mathbf E $ and $\mathbf B $. Since there exists a rotational symmetry of the dispersion of each semimetallic node within the $xy$-plane, the exact directions of $\mathbf E $ and $\mathbf B $ will not matter. What matters is the angle between $ \hat{\mathbf e}_E$ and $ \hat{\mathbf e}_B$. Hence, without any loss of generality, we choose $ \hat{\mathbf e}_E = {\mathbf{\hat x}}  $ and $\hat{\mathbf e}_B = \cos \theta \, {\mathbf{\hat x}} + \sin \theta \, {\mathbf{\hat y}}  $.

\subsubsection{LMC}

The full expression for the LMC is given by
\begin{align}
\sigma_{xx} &= 
\sigma_{xx}^{(1)}   + \sigma_{xx}^{(2)}  +   \sigma_{ xx } ^{(3)}   + \sigma_{ xx } ^{(4)} \,, 
\end{align}
where $\sigma_{xx}^{(1)} $ has already been shown in Eq.~\eqref{eqsxxI1}, and
\begin{align}
\label{C1sigmaxx_bc}
\sigma_{xx}^{(2)}  &= 
  \frac{J^3 \, e^4 \,\tau  \,\alpha^{2/J}  \, B^2 \cos^2 \theta \,
  \, \Gamma (2-1/J ) \, v_0 }
  {16 \,\pi^{ 3/2 } } \, 
  _2\tilde{F}_1 (- 1/J,-1/J - 1/2 ; 5/2- 1/J;\eta^2 )
  \,  \Upsilon_{- 2/J } ( \mu, \beta ) \,,   \nn
  \sigma_{ xx } ^{(3)}  &= \sigma_{ xx } ^{(4)}  = -
   \frac{ J^3 \,e^4 \,\tau \, \alpha^{ 2/J } \, B^2 \cos^2 \theta \,\,
   \Gamma (3- 1/J ) \, v_0 }
 { 32 \, \pi^{ 3/2 } } \, 
 _2\tilde{F}_1 ( 1/2-1/J, -1/J ; 7/2 - 1/J;\eta^2 ) 
 \,\Upsilon_{- 2/J } ( \mu, \beta  )\,.
\end{align}
From Eq.~\eqref{eqdrude}, we find that the Drude part is an even function of $\eta$. The remaining parts, dependent on a nonzero BC, are functions of $\eta^2$, thus making them even functions of $\eta$ as well. For this set-up, only even powers of $B$ appear, and we do not obtain any part linearly varying with $B$, similar to the untilted cases \cite{nag21_magneto, yadav23_magneto}. Also important to note is that the BC-dependent part is independent of the chirality of the node. This is because the terms in the integrand contributing to a nonzero result involve only even powers of the components of the BC (which is proportional to $\chi $). As a consequence of all these observations, we conclude that, if we sum over the contributions from a pair of conjugate nodes, they add up, irrespective of whether of the two nodes are tilted in the same or opposite directions. Fig.~\ref{fig_C1sigmaxx} illustrates the behaviour of the LMC for some representative parameter regimes.

\begin{figure}[t]
\centering
\subfigure[]{\includegraphics[width=0.99 \textwidth]{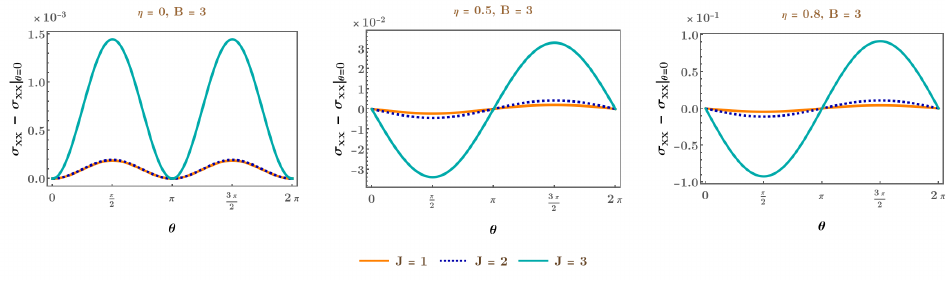}} 
\subfigure[]{\includegraphics[width=0.99 \textwidth]{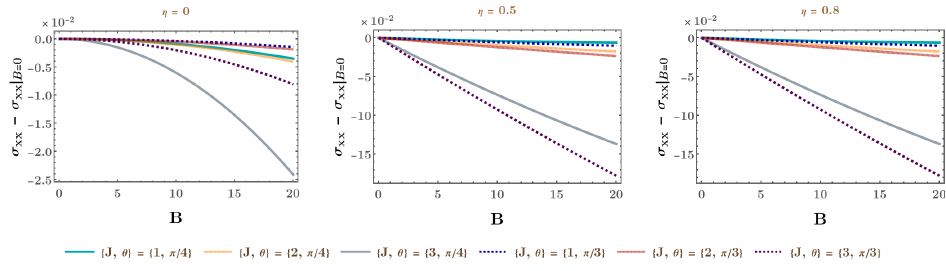}} 
\caption{Set-up II: Variation of the LMC (in units of eV) with (a) the angle $\theta$ and (b) the magnitude $ B $ (in units of eV$^2$) of the magnetic field. For all the plots, we have set $v_0 = 0.005$, $\tau=151$ eV$^{-1}$, $\mu = 0.1$ eV, and $\beta=1160$ eV$^{-1}$.
\label{fig_C2sigmaxx}}
\end{figure}

\begin{figure}[t]
\centering
\subfigure[]{\includegraphics[width=0.99 \textwidth]{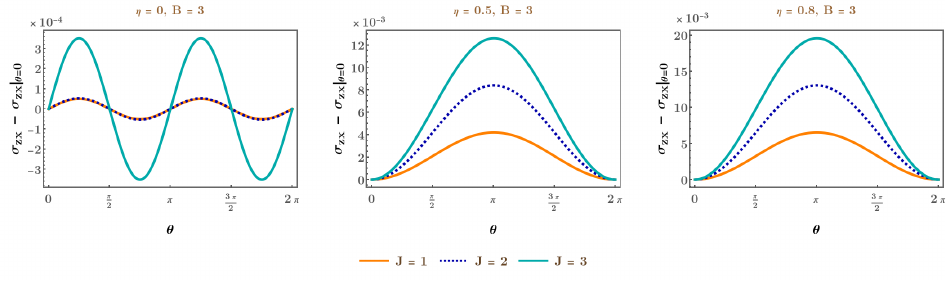}} 
\subfigure[]{\includegraphics[width=0.99 \textwidth]{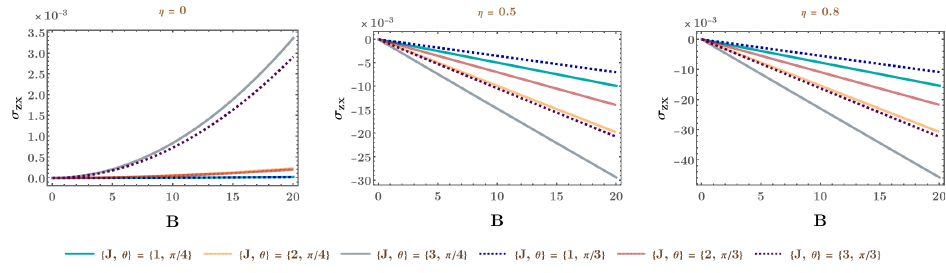}} 
\caption{Set-up II: Variation of the PHC (in units of eV) with (a) the angle $\theta$ and (b) the magnitude $ B $ (in units of eV$^2$) of the magnetic field. For all the plots, we have set $v_0 = 0.005$, $\tau=151$ eV$^{-1}$, $\mu = 0.1$ eV, and $\beta=1160$ eV$^{-1}$.
\label{fig_C2sigmazx}}
\end{figure}

\subsubsection{PHC}

The PHC does not have a Drude part and is entirely caused by a nonzero BC. Explicitly, it takes the form:
\begin{align} 
\label{C1sigmayx}
\sigma_{ yx }  & =
\sigma_{ yx }^{(1)}  + \sigma_{ yx }^{(2)}  + \sigma_{ yx }^{(3)}  + \sigma_{ yx }^{(4)}  \,,
\text{ where}\nn
 \sigma_{ yx }^{(1)} &=  
 \frac{ J^3 \,e^4 \,\tau  \,\alpha^{2/J}\, B^2 \sin (2\theta) \,
 \Gamma (4- 1/J) \,v_0 }   { 128 \, \pi^{ 3/2 } } \,
 _2\tilde{F}_1 ( 1/2-1/J,(J-1)/ J;9/2-1/J;\eta^2 ) 
  \, \Upsilon_{- 2/J } ( \mu, \beta  )\,, \nn
\sigma_{ yx } ^{(2)}  &= 
     \frac{ J^3\, e^4 \,\tau 
 \,   \alpha^{2/J} \,B^2 \sin (2\theta) \, \Gamma (2- 1/J) \, v_0 }
  { 32 \, \pi^{ 3/2 } } \, 
 _2\tilde{F}_1 (- 1/J,-1/J - 1/2; 5/2- 1/J;\eta^2 ) 
      \, \Upsilon_{- 2/J } ( \mu, \beta  )\, , \nn
 \sigma_{ yx }^{(3)}  &=  \sigma_{ yx }^{(4)}  =  
   -\frac{ J^3 \,e^4 \,\tau  \, \alpha^{2/J} 
 \,  B^2 \sin (2\theta) \, \Gamma (3- 1/J) \,v_0 }
 { 64 \, \pi^{ 3/2 } } 
 \, _2\tilde{F}_1 ( 1/2-1/J, -1/J; 7/2- 1/J;\eta^2 )  
 \, \Upsilon_{- 2/J } ( \mu, \beta  )\,.
\end{align}
$\sigma_{ yx }$ is a function of $\eta^2 $ and $B^2 $, and it arises entirely from even powers of the components of the BC (making it independent of the chirality of the node).
Hence, analogous to the PHC, it will give the same contribution from each conjugate node, irrespective of the sign of the tilt. However, unlike the LMC, the PHC vanishes when $\theta$ equals zero or $\pi/2$. Fig.~\ref{fig_C1sigmayx} demonstrates the characteristics of $\sigma_{ yx }$ for some representative parameter regimes.

The PHC components do not show discernible changes with $\beta$, while the LMC does to some extent. Hence, in Fig.~\ref{fig_beta}, we illustrate the variations of LMC with $\beta$.

\subsection{Set-up II} 
\label{eqset2}

In set-up II, as shown in Fig.~\ref{fig_setups}(b), the tilt-axis is perpendicular to $\mathbf E $, but not to $\mathbf B $. We choose $ \hat{\mathbf e}_E =  {\mathbf{\hat x}}  $ and $\hat{\mathbf e}_B = \cos \theta \, {\mathbf{\hat x}} 
+ \sin \theta \, {\mathbf{\hat z}}  $.

\subsubsection{LMC}

The full expression for the LMC is given by
\begin{align}
\sigma_{xx} &= 
\sigma_{\text{Drude}} +\sigma_{xx}^{(1, \text{BC})}  + \sigma_{xx} ^{(2)}  
+   \sigma_{ xx } ^{(3)}   + \sigma_{ xx } ^{(4)} \,, 
\end{align}
where $ \sigma_{\text{Drude}}$ is the same as shown in Eq.~\eqref{eqdrude},
\begin{align}
\label{C2sigmaxx_bc}
\sigma_{xx}^{(1 ,\text{BC})} &
= - \frac{\eta  \,J^2 \,e^3 \, \tau \,(3\, J- 2) \,
 \alpha^{ 2/J } B \sin \theta   \, 
  \Gamma (3- 1/J )  }
 {32 \,\pi^{ 3/2 } \, v_0  } 
 \, _2\tilde{F}_1 (2- 1/J , 5/2- 1/J ; 9/2- 1/J ;\eta^2 ) 
 \, \Upsilon_{2- 2/J } ( \mu, \beta ) 
\nn & \quad 
+ \frac{3 \,J^3\, e^4 \,\tau \,  \alpha^{ 2/J }\, B^2\cos^2 \theta \,
 \, \Gamma (4- 1/J) \, v_0 }
{128 \,\pi^{ 3/2 } }
\, _2\tilde{F}_1 ( 1/2- 1/J , (J-1) / J ; 9/2- 1/J ;\eta^2 ) 
\,\Upsilon_{- 2/J }(\mu, \beta) 
 \nn & \quad  
  + \frac{J^5\, e^4 \, \tau \, \alpha^{\frac{4}{J}} \,B^2 \sin^2 \theta \,
\,\Gamma (4- 2/J )   }
  {64 \,\pi  \, v_0 }  \, 
  _3\tilde{F}_2 (3/2 ,2- 2/J , 3/2- 2/J ; 1/2, 11/2- 2/J ;\eta^2 )
  \, \Upsilon_{2-\frac{4}{J}} (\mu, \beta )\,, 
\end{align} 
\begin{align}
 \sigma_{xx}^{(2)}  & = 
     \frac{ J^3 \,e^4 \,\tau \,  \alpha ^{ 2/J }  \,B^2\cos^2 \theta \,\, 
 \Gamma (2- 1/J )  \, v_0   }
   {16 \,\pi^{ 3/2 } } 
   \, _2\tilde{F}_1 (- 1/J ,-(J+2)/(2 J); 5/2- 1/J ;\eta ^2 )
  \,  \Upsilon_{- 2/J } (\mu, \beta )\,,
\end{align}  
and
\begin{align}   
     \sigma_{ xx } ^{(3)}  &= 
     \sigma_{ xx } ^{(4)}  
  = -\frac{ J^3 \,e^4 \,\tau \,  \alpha^{ 2/J } \, B^2\cos^2 \theta \,\,
  \Gamma  (3- 1/J  )  \, v_0 }
  {32\, \pi^{ 3/2 } }
 \,_2\tilde{F}_1 ( 1/2- 1/J ,- 1/J ; 7/2- 1/J ;\eta^2)
   \, \Upsilon_{- 2/J } (\mu, \beta)\,.
\end{align}
The parts other than the Drude contribution originate from a nonzero BC. In this case, we find that $\sigma_{xx}^{(1 ,\text{BC})}$ has a part which varies linearly in $\eta$ as well as $B$, that originates from the contribution of a term in the integrand which is proportional to the BC (and, consequently, the chirality $\chi $ of the node). This part, being proportional to $\eta \, \chi$, will cancel out when summed over the two nodes in a conjugate pair, if the tilt-signs are the same. The remaining terms are quadratic in both $\eta$ and $B$, and are independent of the chirality. Hence, these parts add up irrespective of the sign of the tilt. Fig.~\ref{fig_C2sigmaxx} illustrates the behaviour of the LMC for some representative parameter regimes. In particular, the periodicity of the curves, as functions of $\theta$, changes from $\pi$ to $2\pi $ as soon as a nonzero tilt is introduced. Furthermore, Fig.~\ref{fig_C2sigmaxx} (b) shows that the linear-in-$B$ parts dominate for a nonzero tilt, with the magnitude increasing with increasing $J$ [because of the factor $ J^2 \, (3\,J-2)$, which monotonically increases with $J$]. Most importantly, the dominant $B$-linear terms lead to a change in the overall sign of the LMC for various ranges of $\theta$, when measured with respect to the $B=0$ case. We also capture the $\eta$-dependence in Fig.~\ref{fig_eta}.

Let us elaborate a bit more on the appearance of the term proportional to $\eta\,B\,\chi $. When the system is subjected to homogeneous external fields, then, in the absence of any other scale in the problem, the Onsager-Casimir reciprocity
relation $\sigma_{xx} (\mathbf{B})= \sigma_{xx} (-\mathbf{B})$ \cite{onsager31_reciprocal, onsager2, onsager3} forbids any term in the LMC to be linear in $B$, unless the change of sign of $\mathbf B $ is compensated by a change of
sign in another parameter in the system. The tilt vector, defined by $\mathbf t = \eta \, {\mathbf{\hat z}}$ in this paper, provides us with such a compensating sign, such that we have now the identity $\sigma_{xx} (\mathbf{B}, \mathbf t)= \sigma_{xx} (-\mathbf{B}, -\mathbf t )$, thus fulfilling the Onsager-Casimir constraints~\cite{cortijo16_linear}. This allows the linear-in-$B$ to appear in $\sigma_{xx}^{(1 ,\text{BC})}$, with the corresponding part of the magnetocurrent being proportional to $ \left ( \mathbf t \cdot \mathbf B \right )  \hat{\mathbf e}_E $, in agreement with the findings of Ref.~\cite{das19_linear} for WSMs at $T = 0 $. Because $ \hat{\mathbf e}_E$ is perpendicular to $\mathbf t $ in this set-up, the linear contribution vanishes for $\theta = 0 $. We note that this term has a nontrivial dependence on the chemical potential and the temperature, via the term $\Upsilon_{ 2- 2/J } (\mu, \beta)$, for $J \neq 1$ (because, of course, $\Upsilon_0 = 1$). The fact that the $\mu$-dependence disappears for $J=1$ is consistent with the WSM studies of Ref.~\cite{das19_linear}.

\begin{figure}[t]
\centering
\centering
\subfigure[]{\includegraphics[width=0.99 \textwidth]{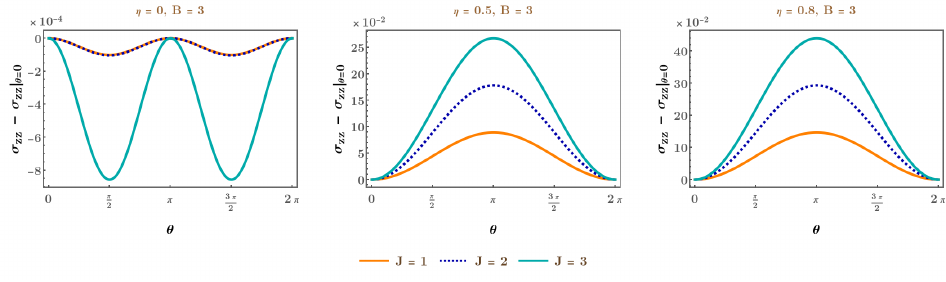}} 
\subfigure[]{\includegraphics[width=0.99 \textwidth]{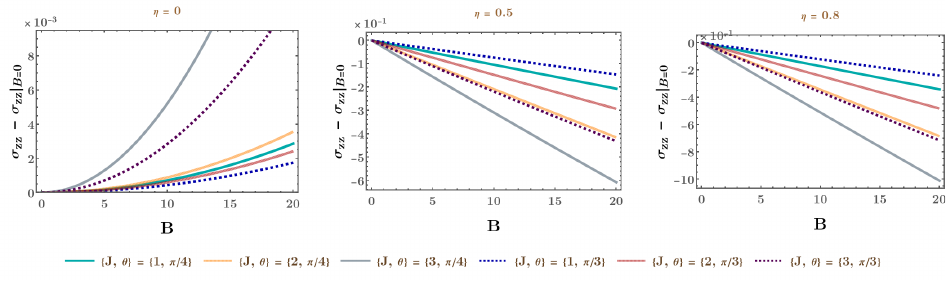}} 
\caption{Set-up III: Variation of the LMC (in units of eV) with (a) the angle $\theta$ and (b) the magnitude $ B $ (in units of eV$^2$) of the magnetic field. For all the plots, we have set $v_0 = 0.005$, $\tau=151$ eV$^{-1}$, $\mu = 0.1$ eV, and $\beta=1160$ eV$^{-1}$.
\label{fig_C3sigmazz}}
\end{figure}

\subsubsection{PHC}

\begin{figure}[t]
\centering
{\includegraphics[width=0.99 \textwidth]{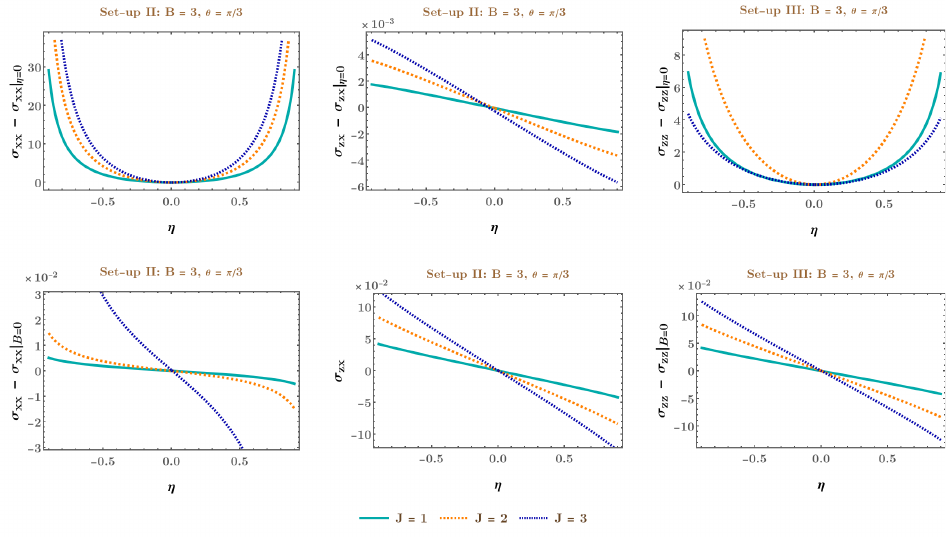}} 
\caption{Comparison of the $\eta$-dependence of the $\sigma_{pq}$ components (in units of eV) for the two set-ups (indicated in the plotlabels) in which we find linear-in-$B$ terms.
For all the plots, we have set $ B = 3$ eV$^2$, $\theta = \pi/3$, $v_0 = 0.005$, $\tau=151$ eV$^{-1}$, $\mu = 0.1$ eV, and $\beta=1160$ eV$^{-1}$. The curves in the upper(lower) panel have the $\eta=0$($B=0$) parts subtracted off from the full response.
\label{fig_eta}}
\end{figure}

Analogous to set-up I, the PHC does not have a Drude part and is entirely caused by a nonzero BC. Explicitly, it takes the form:
\begin{align} 
\sigma_{ zx }   =
\sigma_{ zx }^{(1)}  + \sigma_{ zx }^{(2)}  + \sigma_{ zx }^{(3)}  + \sigma_{ zx }^{(4)}  \,,
\nonumber
\end{align}
where
\begin{align} 
\label{C2sigmazx}
\sigma_{zx}^{(1)}  & = 
- \frac{ J\, e^3\, \tau  
\left[ 3 \,\eta  - 5\, \eta^2 - 3 \left(\eta^2-1\right)^2
 \tanh^{-1}\eta \right]
B \cos \theta \,v_0 \left  (1-\delta_{\eta,0}  \right )   
}
{24 \,\pi^2  \,\eta^4  } 
\nn & \quad
 + 
\left[ \sqrt{\pi } \,J \,\,
  _3\tilde{F}_2 ( 3/2, (J-1) / J, 1/2- 1/J; 1/2, 9/2- 1/J;\eta^2 )    
   -\eta^2 \left (J-2 \right ) 
 \, _2 \tilde{F}_1 ( 3/2- 1/J, (J-1) / J; 9/2- 1/J;\eta^2 )
\right] \nn
&  \hspace{1 cm } \times \frac{
J^2 \,e^4\, \tau \, \alpha^{2/J} \,B^2\sin (2\theta )\,
\, \Gamma(3-{1}/{J})\, v_0 \,  \Upsilon_{-2/J} ( \mu, \beta ) 
}
 { 64 \, \pi^{ 3/2 } } \,,
 \end{align}
\begin{align}
\sigma_{zx}^{(2)}  & = 
\frac{ J^3 \,e^4 \,\tau \, \alpha^{ 2/J}  \,
B^2 \sin (2\theta ) \,\Gamma  (2- 1/J) \, v_0 }
{ 32 \, \pi^{ 3/2 } } \, \,
_2\tilde{F}_1 (- 1/J,-(J+2)/(2 \,J); 5/2- 1/J;\eta^2 )
\, \Upsilon_{- 2/J} ( \mu, \beta )\,,
\end{align}
\begin{align}
\sigma_{zx}^{(3)}  & =
 -\frac{J^3 \,e^4 \,\tau \, \alpha^{ 2/J}  \, B^2 \sin (2\theta ) \,
 \Gamma (3- 1/J )\, v_0 }
 { 64\, \pi^{ 3/2 } } 
 \, \, _2\tilde{F}_1 ( 1/2 - 1/J,- 1/J; 7/2 - 1/J;\eta^2 ) \,
 \Upsilon_{- 2/J} ( \mu, \beta )\,,
 \end{align} 
and
\begin{align} 
 \sigma_{zx}^{(4)}  & =
  -\frac{\eta \, J\, e^3\, \tau  \, B \cos \theta \,v_0   }
  {4 \,\pi^2  }   
 \nn &  \qquad  -
 \left[ \sqrt{\pi } \,J \, _3\tilde{F}_2 ( 3/2 , 1/2 - 1/J,- 1/J; 1/2 , 7/2 - 1/J;\eta^2 )  
 + 2\, \eta^2 \, _2\tilde{F}_1 ( 1/2 - 1/J,(J-1)/{J}; 7/2 - 1/J;\eta^2 )
\right ] \nn & \hspace{ 1.5 cm }
\times 
\frac{ J^2\, e^4 \,\tau \, \alpha^{ 2/J} \,
B^2 \sin (2\theta )  \,
 \Gamma (2- 1/J)\, v_0 \,  \Upsilon_{- 2/J} ( \mu, \beta )}
 { 64 \, \pi^{ 3/2 } } \,.
\end{align}
Here too we find the emergence of terms proportional to $\text{sgn}(\eta) \,B\,\chi $ [cf. the first terms in $\sigma_{zx}^{(1)}$ and  $\sigma_{zx}^{(4)}$], but without having any dependence on $\mu$ or $ \beta $ for any $J$ (unlike the LMC). Moreover, they are directly proportional to $J$ as well.
The corresponding part of the magnetocurrent is proportional to $ \left ( \hat{\mathbf e}_E \cdot \mathbf B \right ) 
\mathbf t  $, in agreement with the findings of Ref.~\cite{das19_linear} for WSMs at $T = 0 $. Therefore, this linear contribution vanishes when $\theta =\pi/2$. Fig.~\ref{fig_C2sigmazx} demonstrates the characteristics of the full PHC for some representative parameter regimes, where we find the linear term to be dominating the overall behaviour of the curves for tilted nodes, with the magnitude increasing with increasing values of $J$. A nonzero tilt also changes the periodicity of the curves, as functions of $\theta$, from $\pi$ to $2\pi $, and causes an overall change in the sign of the PHC
for various ranges of $\theta$, when measured with respect to the $B=0$ case.  We depict the $\eta$-dependence in Fig.~\ref{fig_eta}.

The PHC components do not show discernible changes with $\beta$, while the LMC does to some extent. Hence, in Fig.~\ref{fig_beta}, we illustrate the variations of LMC with $\beta$.

\subsection{Set-up III}

In set-up III, as shown in Fig.~\ref{fig_setups}(c), the tilt-axis is parallel to $\mathbf E $, such that $ \hat{\mathbf e}_E =  {\mathbf{\hat z}}  $, and $\hat{\mathbf e}_B = \cos \theta \, {\mathbf{\hat z}} + \sin \theta \, {\mathbf{\hat x}}  $.

\subsubsection{LMC}

The full expression for the LMC is given by
\begin{align}
\sigma_{zz} &= 
\tilde \sigma_{\text{Drude}} +\sigma_{zz}^{(1, \text{BC})}  + \sigma_{zz} ^{(2)}  
+   \sigma_{zz}^{ (3)}   + \sigma_{zz} ^{(4)} \,, 
\end{align}
where
\begin{align} 
\tilde \sigma^{\text{Drude}} &=  
\frac{ e^2 \, \tau  \,v_0 \Upsilon_{2/J}  ( \mu, \beta ) }
{ 8 \, \pi^{  3/2 } \,\alpha^{2/J}  \, \Gamma  ( 3/2+ 1/J)
}
\left[ 
\Gamma (1+ 1/J )\, \delta_{\eta, 0} 
+  
\frac{  J + 2 -2 
\, \, _2F_1 (  1/2 ,1;  3/2 + 1/J ;\eta^2)
}
{ \left(1-\eta^2\right)^{1/J}   \,J^2 } 
\, \Gamma ( 1/J ) \, (1 - \delta_{\eta, 0}) 
\right] ,
\end{align}
\begin{align}   
\label{C3sigmazz}
 & \sigma_{zz}^{ (1, \text{BC}) } 
\nn  & = -\frac{
  J \, e^3 \,\tau \, B \cos \theta \, v_0
 \left[ 3 \left(\eta^2-1\right)^2 \tanh^{-1} \eta 
- 3\, \eta^5 + 5 \,\eta^3 - 3 \,\eta \right ]
\left( 1-\delta_{\eta,0} \right )
}
{12\, \pi^2 \, \eta^4  } 
\nn & \quad + 
\frac{ 
J \, e^4 \,\tau \,  B^2 \cos^2 \theta \, v_0^3
\left(7 \,\eta^2 + 1\right)  \Upsilon_{-2} ( \mu, \beta ) 
}
{120\, \pi^2 }     
 \nn &  
\quad + 
\Bigg[ 2\, \eta^2 J \, 
 _3\tilde{F}_2 (  3/2 ,(J-1) / {J},  1/2 - 1/J ;  1/2 ,7/2- 1/J ;\eta^2)  
 +  3 \,J \, 
    _3\tilde{F}_2 (5/2 ,(J-1) / {J},  1/2 - 1/J ;  1/2 ,9/2- 1/J ;\eta^2 ) \nn
&  \hspace{1 cm}
-3\, \eta^2 \left (J-2 \right ) \, 
\,
_3\tilde{F}_2\left( 5/2 ,(J-1) / {J},  3/2 - 1/J ;  3/2 ,9/2- 1/J ;\eta^2\right)    
\Bigg ]  
 \frac{J^2\, e^4 \,\tau  \, \alpha^{ 2 / J }  \, B^2 \cos^2 \theta \,
  \Gamma (2-1/J ) \, v_0
\, \Upsilon_{-2/J}  ( \mu, \beta )   }   
 { 64 \,\pi  } 
\, ,  
\end{align}

\begin{figure}[t]
\centering
{\includegraphics[width=0.99 \textwidth]{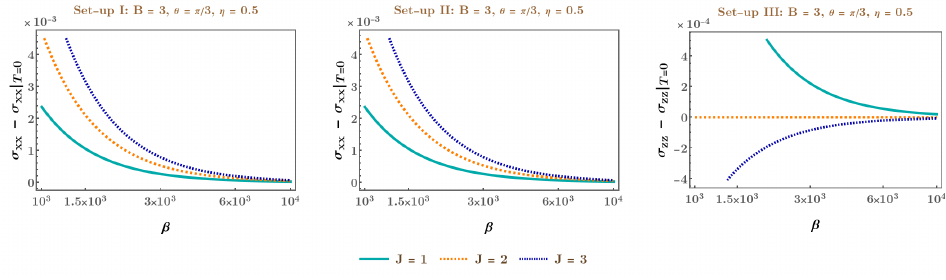}} 
\caption{Comparison of the variations of the LMC components (in units of eV) as functions of $\beta$ (in units of eV$^{-1}$) over various set-ups (indicated in the plotlabels). For all the plots, we have set $ B = 3$ eV$^2$, $\theta = \pi/3$, $v_0 = 0.005$, $\tau=151$ eV$^{-1}$, $\mu = 0.1$ eV, and $\eta =0.5 $. The curves have the $ T =0$ parts subtracted off from the full response.
\label{fig_beta}}
\end{figure}

\begin{align}
     \sigma_{zz}^{ (2) } &=  
  \frac{ J^3 e^4 \tau  \alpha^{ 2 / J }  
  \, B^2 \cos^2 \theta \,  \,\Gamma (2-1/J ) \, v_0
\, \Upsilon_{-2/J}  ( \mu, \beta )   }
 {16 \,\pi^{  3/2 } } \, _2\tilde{F}_1 \big (- 1/J ,-(J+2)/(2 J); 5/2 - 1/J ;\eta^2 \big )  \, , 
\end{align}
\begin{align}      
     \sigma_{zz}^{ (3) } &   = \sigma_{zz}^{ (4) } 
 =  \frac{ \eta\,  J \,e^3 \,\tau \,B \cos \theta \, v_0  }  
  {4 \, \pi^2  } 
 \nn &
\hspace{ 1.5 cm }  + 
 \Bigg[ \sqrt{\pi } 
\, J \,\, 
_3\tilde{F}_2 (  3/2 ,  1/2 - 1/J ,- 1/J ;  1/2 , 7/2- 1/J ;\eta^2 )      
+ 2 \,\eta^2 \, _2\tilde{F}_1 (  1/2 - 1/J ,(J-1) /{J}; 7/2- 1/J ;\eta^2 )
  \Bigg ] \nn & \hspace{ 2.2 cm} \times
 \frac{ J^2\, e^4 \,\tau \, \alpha^{ 2 / J }  \,  B^2 \cos^2 \theta\,   \Gamma (2-1/J ) \, v_0
  \, \Upsilon_{-2/J}  ( \mu, \beta ) }
  {32\, \pi^{  3/2 } }  \, .
\end{align}
We find that there is a $B$-independent Drude part, viz., $ \tilde \sigma^{\text{Drude}} $, whose form is different from Eq.~\eqref{eqdrude}. However, similar to $  \sigma^{\text{Drude}} $, $ \tilde \sigma^{\text{Drude}} $ is an even function of $\eta $ (independent of $\chi $) and, hence, independent of the sign of the tilt.  Analogous to the LMC in set-up II, we find that $\sigma_{zz}^{ (1, \text{BC}) } $ contains a linear-in-$B$ term, whose behaviour goes as $\propto \text{sgn} (\eta) \, \chi \, B \, J$. In addition, there is another linear-in-$B$ term coming from $\sigma_{zz}^{ (3) }$ and $ \sigma_{zz}^{ (3) }$ with the dependence $\propto \eta \,B\,\chi \, J $. None of these $B$-linear terms has any $\mu$- or $T$-dependence and, while summing over the conjugate nodes of opposite chiralities, the net result will be nonzero only for tilting in opposite directions. The part of the electric current arising from them is proportional to $ \left ( \mathbf t \cdot \mathbf B \right )  \hat{\mathbf e}_E $, and the identity $\sigma_{zz} (\mathbf{B}, \mathbf t)= \sigma_{zz} (-\mathbf{B}, -\mathbf t )$ makes it possible to satisfy the Onsager-Casimir reciprocity relations.
The remaining terms of $ \sigma_{zz}$ are quadratic in $B$, contain even powers of $\eta$, and are independent of the chirality of the node. All the non-Drude terms  vanish when $\theta = \pi/2$ and have an overall $2\pi$($\pi$)-periodicity in $\theta $ for nonzero(zero) $\eta $. All these are accompanied by an overall change in sign of the LMC for various ranges of $\theta$, when measured with respect to the $B=0$ case. The typical characteristics of the LMC are captured in Fig.~\ref{fig_C3sigmazz} via some representative parameter regimes. We also capture the $\eta$-dependence in Fig.~\ref{fig_eta}. Fig.~\ref{fig_beta} illustrates the variations of the LMC with $\beta$.

\subsubsection{PHC}

As for the PHC, we have the simple relation
\begin{align}
\sigma_{xz}( \theta ) \big \vert _{\text{set-up III}}= 
\sigma_{zx}( \pi/2 - \theta )
\big \vert _{\text{set-up II}} \,.
\end{align} 
Hence, we get a linear term similar to set-up II. The only difference is that here the magnetocurrent, arising from the $B$-linear term, is proportional to $ \left ( \hat{\mathbf e}_E \cdot \mathbf t \right ) 
B \sin \theta \, {\mathbf{\hat x}} $.

\subsection{Other set-ups and generic conclusions}

Let us consider another distinct set-up where we have $ \hat{\mathbf e}_E = {\mathbf{\hat z}} $ and $ \hat{\mathbf e}_B = \cos \Phi \,{\mathbf{\hat x}} +  \sin \Phi \, {\mathbf{\hat y}} $. This is not a planar Hall set-up, as $\mathbf E$ and $\mathbf B$ are perpendicular to each other, and, hence, only gives rise to the transverse (or Hall) conductivity. This also implies that the Drude part has to be zero. 
The PHC components are explicitly given by
\begin{align} 
\sigma_{ xz }  &  =
\sigma_{ xz }^{(1)}  + \sigma_{ xz }^{(2)}  
+ \sigma_{ xz }^{(3)}  + \sigma_{ xz }^{(4)}  \,, \nn
\sigma_{yz}(\Phi) & = \sigma_{xz}(\pi/2 -\Phi) 
\text{ (because of the symmetric dispersion in the $xy$-plane)}\,,
\end{align}
where
\begin{align}  
\label{C4sigmaxz BC}
\sigma_{xz}^{^{(1)}} 
&	= -\eta\, B  \cos \Phi   \left [ 
\frac{ J \,e^3\,v_0   \left \lbrace  3\, \eta  - 5\, \eta^2 
+ 3 \left( 1- \eta^2 \right)^2 \tanh^{-1} \eta \right \rbrace 
\sin \Phi  \, {\mathbf{\hat z}} }
{24 \, \pi^2 \, \eta^4 
 } 
 + \order{B^2} 
\right ] ,\nn
\sigma_{xz}^{^{(2)}} & = \sigma_{xz}^{^{(3)}} =  0\,, \quad
\sigma_{xz}^{^{(4)}} = 
- \eta \,B \cos \Phi \left[  \frac{ J \, e^3 \, \tau \, v_0}
{4\, \pi^2 } + \order{B^2} \right]  .
\end{align}
The $B^2$-dependent terms are identically zero in this case and, to any order in $B$, the final expression is proportional to the tilt $\eta$. This results from the fact that only odd powers of $B$ exist upto any order.

The components of $\sigma_{pq}$ in all other possible configurations of PHE set-ups can be obtained from the ones shown here. For example, if we take $\hat{\mathbf e}_E  = {\mathbf{\hat x}}$ and $\hat{\mathbf e}_B = \cos \theta \, {\mathbf{\hat x}} 
+ \sin \theta \left( \cos \Phi  \, {\mathbf{\hat z}} + \sin \Phi  \, {\mathbf{\hat y}} \right)  $, which is the generalization of set-up II with $\mathbf B $ now having a nonzero $y$-component as well, the extra PHC component $\sigma_{yx}$ is the same as that obtained for set-up I, after making the replacement $\sin \theta \rightarrow  \sin \theta \, \sin \Phi $. Another example is when we choose $ \hat{\mathbf e}_E  = {\mathbf{\hat x}} $ and $ \hat{\mathbf e}_B = \cos \Phi  \,{\mathbf{\hat y}}  + \sin \Phi \, {\mathbf{\hat z}} $ --- the corresponding PHC expressions are:
\begin{align}
  \sigma_{yx} = \sigma_{yx}(\theta= {\pi} /{2}) \big \vert_{\text{set-up I}} = 0\,,
 \quad 
  \sigma_{zx} = \sigma_{zx} (\theta= {\pi} /{2}) \big \vert_{\text{set-up II}} = 0\,.\nonumber
\end{align}
Hence, our results exhaust all the distinct cases that can arise, taking into account all possible relative orientations and direction-dependence.

\section{Summary and outlook}
\label{secsum}

The PHE has been measured in numerous experiments --- a few examples involve materials like ZrTe$_{5}$ \cite{li18_giant, li_2016}, TaAs \cite{checng-long}, NbP \cite{li_nmr17}, NbAs \cite{li_nmr17}, and Co$_3$Sn$_2$S$_2$ \cite{shama}, hosting Dirac and Weyl nodes. In particular, signatures of linear-in-$B$ behaviour have been reported in their data.
Ref.~\cite{huang15_observation} has presented experimental evidence confirming the giant LMR resulting from the chiral anomaly in the time-reversal-invariant Weyl semimetal material $\mathrm{TaAs}$, and has shown that the response undergoes a sign change as the direction of the magnetic field is rotated. 
In Ref.~\cite{li18_giant}, by considering $\mathrm{ZrTe_{5-\delta}}$, the authors have shown the variation of the planar Hall resistivity with the angular orientation of the applied magnetic field. The experimental results in Ref.~\cite{claudia_phe} show the angular variations of both the longitudinal and the planar Hall resistivity components in the Heusler Weyl semimetal $\mathrm{GdPtBi}$. The measurements reported in Ref.~\cite{wu18_probing} show the presence of both linear-in-$B$ and $B^2$-dependent terms in the response for Cd$_3$As$_2$.
In order to find the origins of the experimental data, as well as to motivate further investigations, we have delved into the characterization of the magnetoelectric conductivity tensors in the context of planar Hall set-ups built with tilted WSMs and mWSMs, considering various relative orientations of the electromagnetic fields and the direction of the tilt.

One instance of the importance of our studies can be motivated as follows.
We know that the response in PHE is not solely generated by the chiral anomaly and nonzero BC in the nodal-point semimetals. In fact, it can also have nontopological origins from magnetic order, spin-orbital coupling, and trivial in-plane orbital magnetoresistance \cite{nontopo_phe}. The PHE in nonmagnetic
WSMs/mWSMs is dominated by the BC and chiral anomaly contributions only for materials with a relatively small orbital magnetoresistance \cite{claudia_phe}. However, in magnetic WSMs/mWSMs, the response is dominated by the part originating from the Berry phase, because the anisotropic magnetic resistance (caused by different in-plane and out-of-plane spin scatterings) induced in ferromagnetism is very small in comparison. The topological and nontopological origins of the PHE, discussed above, cannot always be clearly distinguished in experiments, because the experimentally observed quadratic-in-$ B $ magnetoconductance arises via both mechanisms. To remedy the situation, we can use the fact that the BC gives rise to a linear-in-$ B$ term in magnetoconductance if the node is tilted. In fact, our results show that this term dominates in suitable parameter regimes, and increases in magnitude as the magnitude of the monopole charge $J$ increases. Most importantly, the dominant $B$-linear terms lead to a change in sign of the magnetoconductance depending on the value of $\theta$, when measured with respect to the $B=0$ case.

If we consider a 3d Dirac semimetal with the Hamiltonian~\cite{yang_dirac_weyl}
\begin{align}
\label{eqdirac}
	\mathcal{H}_{\text{dirac}} =
 k_x \,\sigma_y \, \tau_x  - k_y \,\sigma_x \,\tau_x  
+  k_z \,\tau_z \,,
\end{align}
where $\boldsymbol{\sigma}$ and $\boldsymbol{\tau}$ denote the vectors of the three Pauli matrices acting on two different symmetry-operator-spaces, we find that the Dirac node comprises a pair of doubly-degenerate cones with dispersions $  \pm k$. Adding a perturbation of the form $\delta \, \sigma_x \, \tau_y$ splits the Dirac node into two distinct Weyl nodes, separated in the $k_x$-momentum direction. A single Weyl node has the disperion relation $\pm E_{\text{weyl},s}$, where $E_{\text{weyl},s} =  \sqrt{ \left [ k_x + (-1)^s\, \delta \right ]^2+k_y^2+k_z^2}$, and $s=1,2$ labels the two Weyl nodes. 
On linearizing the momentum dependence around each node, the problem reduces to a $2\times 2$ Hamiltonian
$\mathcal{H}_\chi = \chi\,\boldsymbol{k} \cdot \boldsymbol{\sigma}$ in the vicinity of each node, where $\chi$ denotes the chirality of the node in consideration. In the experimental results reported in papers like Ref.~\cite{dirac_exp_xiong}, Dirac semimetals such as Na$_3$Bi are considered. There, an applied magnetic field breaks the time-reversal symmetry and separates the two nodes due to the spin Zeeman energy. Now, if an electric is applied parallel to the magnetic field, an axial current is generated, which is observed as a large negative LMR. Hence, our results are applicable to experimental observations on Dirac semimetals, when the node gets split into two Weyl nodes due to an external perturbation (e.g., the applied magnetic field).

We would like to point out that we have not considered the OMM in order to make it possible to obtain closed-form expressions from the integrals and to identify the features (e.g., emergence of the linear-in-$ B $ dependence) originating solely from the effects of tilt. This is justified by the fact that the OMM, in the limit of negligible internode scatterings, is not expected to change the qualitative results, and has been neglected in the earlier studies as well \cite{nandy_2017_chiral, sharma17_chiral, ma19_planar, das19_linear, pal22b_berry}. However, in future we will determine how a combination of OMM and significant internode scatterings modify the response, as indicated in Refs.~\cite{das-agarwal_omm, timm_pos_lmr}. In order to capture the correct physics, we must go beyond the relaxation-time-approximation \cite{timm_pos_lmr}.

In the future, it will be worthwhile to study the influence of a strong quantizing magnetic field, when Landau level formation cannot be ignored \cite{ips-kush, fu22_thermoelectric,staalhammar20_magneto, yadav23_magneto}. The behaviour of the  magnetoconductivity in the presence of strong disorder as well as many-body interactions \cite{ips-seb, ips_cpge, ips-biref, ips-klaus, rahul-sid, ipsita-rahul-qbt, ips-qbt-sc} will take us into the regime where interactions cannot be ignored. Another possible direction is to investigate the response resulting from an additional time-periodic drive \cite{ips-sandip, ips-sandip-sajid, ips-serena} applied on the system (e.g., by shining a beam of circularly polarized light).

\section*{Acknowledgments}
RG is grateful to Ram Ramaswamy for providing the funding during the initial stages of the project. IM's research leading to these results has received funding from the European Union's Horizon 2020 research and innovation
programme under the Marie Skłodowska-Curie grant agreement number 754340.

\bibliography{ref_tilt}

\end{document}